\documentclass[aps,preprint,epsfig,rotate]{revtex4}
\usepackage{graphicx}
\usepackage{bm}
\usepackage{epsfig}

\input epsf
\begin{document}
%\begin{doublespace}

\title{Additional ionization during nuclear $\beta^{\pm}$ decay in atomic systems}       

\author{Alexei M. Frolov}
\email[E--mail address: ]{alex1975frol@gmail.com}

\affiliation{Department of Applied Mathematics \\
 University of Western Ontario, London, Ontario N6H 5B7, Canada}

\date{7 October, 2026}
%\date{\today}

\begin{abstract}

An original method is developed for numerical calculations of probabilities 
of the bound-free atomic excitations during the nuclear $\beta^{\pm}$ decays 
in actual atoms and ions. To achieve this goal we have derived and tested 
the closed analytical formulas for the final state probabilities of 
different excitations in the few-electron atomic systems. This study is the 
first non-approximate and essentially complete investigation of the atomic 
bound-free excitations during the nuclear $\beta^{\pm}$ decays in atoms. 

%\noindent 
%DOI: 10.13140/RG.2.2.29994.38084

\end{abstract}

\maketitle
\newpage

\section{Introduction} 

As is well known (see, e.g., \cite{Blat}) the nuclear $\beta^{\pm}$ decay 
in any atom and/or ion proceeds with the emission of one fast $\beta^{-}$ 
electron (or $\beta^{+}-$positron) and neutrino (or anti-neutrino). In 
the original atom the process of nuclear $\beta^{\pm}$ decay produces a 
sudden change in the central atomic nucleus, where its nuclear charge $Q$ 
increases (or decreases) by unity and becomes $Q \pm 1$. Formally, the 
electron density distribution around central nucleus must be changed too. 
However, such a change in the original electron density always proceeds 
with a certain delay. This well known experimental fact was used to 
develop the so-called 'sudden approximation' \cite{Mig1}, \cite{Mig2}. In 
reality, the original electron density distribution can be used to 
determine the final state probabilities of formation of certain, final 
atomic states. 

Note that the velocity of the emitted $\beta^{\pm}$ electron is usually 
a very large value which is often close to the speed of light in vacuum. 
In fact, such a fast particle penetrates all electron shells of the 
incident atom in a very short time. Direct electron-electron excitations 
of bound atomic electrons during this penetration process can be 
evaluated as $\simeq \alpha^{2}$, where $\alpha \approx \frac{1}{137}$ 
is the fine-structure constant, i.e., it is a negligible value. A unique 
combination of two crucial facts that: (1) the electron density 
distribution in the maternal atom does not change noticeable during the 
nuclear $\beta^{\pm}$ decay, and (2) very small (or negligible) 
probability of the direct electron-electron excitations allows us to use 
the incident electron density distribution in order to evaluate (or 
predict) the final state probabilities in the final atom/ion. Below, we 
shall assume that our main goal is achieved, if we can calculate the 
final state probabilities of various (final) atomic states. 

The both ideology and methodology of the sudden approximation for bound 
state spectra in atoms/ions are well described in our earlier papers 
(see, e.g., \cite{Fro2005} - \cite{Fro2016} and others). The general 
equation of the nuclear $\beta^{-}$-decay in some few-electron atom $X$ 
is written in the following form: 
\begin{equation}
 X(n L S) = Y^{+}(n^{\prime} L S) + \beta^{-} + \nu \; \; \; ,  \; 
\; \label{beta} 
\end{equation}
where the original atom $X$ was in some bound state, while the final ion 
$Y^{+}$ is also in one of its bound states. In this study the notation 
$Y$ stands for the atomic nucleus which contains one positively charged 
proton $p^{+}$ more than the original atomic nucleus of $X$. If one 
simply calculates the overlap integral between the bound state wave 
functions of the original atom and final ion, then its absolute value 
determines the corresponding amplitude ${\cal A}_{if}$. The following 
calculation of the final state probability ${\cal P}_{ij} = \mid {\cal 
A}_{ij} \mid^{2}$ is straightforward and this numerical value can be 
compared with the results, which were measured in actual experiments. 
The sum of these partial probability amplitudes ${\cal P}_{ij}$ over 
all possible final bound states gives us the total final state 
probability of `atomic excitations' during the nuclear $\beta^{\pm}$ 
decay. 

Let us describe the procedure of determining the probability amplitude 
${\cal A}_{if}$ with more details by using a few additional quantum 
notations. The overlap integral of the two bound-state, $N-$electron 
wave functions $\Psi_{X}$ and $\Psi_{Y^{+}}$ is written in the form 
\begin{eqnarray}
 {\cal A}_{if} = \int \Bigl[ \hat{A}_{N} \Psi_{X}({\bf x}_{1}, {\bf x}_{2}, 
 \ldots, {\bf x}_{N})\Bigr] \Bigl[ \hat{A}_{N} \Psi_{Y^{+}}({\bf x}_{1}, 
 {\bf x}_{2}, \ldots, {\bf x}_{N}) \Bigr] d^{3}{\bf x}_{1}d^{3}{\bf x}_{2} 
 \ldots d^{3}{\bf x}_{N} \; \; \; , \; \; \label{overl} 
\end{eqnarray}
where ${\bf x}_{k} = ({\bf r}_{k}, {\bf s}_{k})$ and $k = 1, 2, \ldots N$ 
are the Cartesian coordinates of the $k-$the electron and $\hat{A}_{N}$ 
is the projector operator which performs a complete anti-symmetrization of 
all spatial and spin coordinates for $N$ identical electrons. Everywhere 
below in this study we shall assume that the antisymmetric wave function 
of any bound atomic state are represented in the form of determinants which 
include all `optimal' one-electron spin-orbitals. In particular, the wave 
function of the original neutral atom takes the form:
\begin{eqnarray}
 \Psi_{X}({\bf x}_{1}, {\bf x}_{2}, \ldots, {\bf x}_{N}) = \hat{A}_{N} 
 \Psi_{X} = \frac{1}{\sqrt{N!}} \; Det \mid \psi_{\alpha_i}({\bf r}_{k}, 
 {\bf s}_{k}) \mid \; \; \; , \; \; \label{psi1} 
\end{eqnarray}
where the index $i$ designates the one-electron $\psi_{\alpha_i}({\bf r}_{k}, 
{\bf s}_{k})$ spin-orbitals \cite{McWen} which are occupied by the $k-$th 
electron. Here the index $\alpha_{i}$ indicates the number of row (or index 
of row), while $k$ is the column number (or index of column). It is also 
assumed here that the indices $\alpha_1, \alpha_2, \ldots, \alpha_{N}$ are 
arranged in increasing order. Similar formula can be written for the wave 
function of the final $Y^{+}$ ion  
\begin{eqnarray}
 \Psi_{Y^{+}}({\bf x}_{1}, {\bf x}_{2}, \ldots, {\bf x}_{N}) = \hat{A}_{N} 
 \Psi_{Y^{+}} = \frac{1}{\sqrt{N!}} \; Det \mid \phi_{\beta_i}({\bf r}_{n}, 
 {\bf s}_{n}) \mid \; \; \; , \; \; \label{phi1} 
\end{eqnarray}
where again the index $i$ designates the one-electron $\psi_{\beta_i}({\bf 
r}_{n}, {\bf s}_{n})$ spin-orbitals which are occupied by the $n-$th 
electron, where $1 \le n \le N$, where $1 \le n \le N$. The index $\beta_{i}$ 
indicates the number of row (or index of row) in this (final) determinant, 
while the index $n$ is the column number (or index of column). Note that the 
both wave functions (or determinants), Eqs.(\ref{psi1}) and (\ref{phi1}), 
contain the same number of bound electrons $N$ and this simplifies all 
following calculations. Finally, we have to re-derive the formula, 
Eq.(\ref{overl}), for the probability amplitude by re-writing it in terms of 
the wave functions Eqs.(\ref{psi1}) - (\ref{phi1}), each of which is 
represented as the corresponding determinant. The arising formula is: 
\begin{eqnarray}
 {\cal A}_{if} = \frac{1}{N!} \; \int \; Det \mid \psi_{\alpha_i}({\bf 
 r}_{k}, {\bf s}_{k}) \mid Det \mid \phi_{\beta_i}({\bf r}_{n}, {\bf 
 s}_{n}) \mid d^{3}{\bf x}_{1}d^{3}{\bf x}_{2} \ldots d^{3}{\bf x}_{N} 
 \; \; \; , \; \; \label{psiphi} 
\end{eqnarray}
and it solves the problem for bound-bound atomic excitations during the 
nuclear $\beta^{\pm}$ decays in $N-$electron atoms and ions. It is assumed 
in Eq.(\ref{psiphi}) and everywhere below in this study that $1 \le k \le 
N$ and $1 \le n \le N$. 
 
As mentioned above the numerical value of the overlap integral between the 
original and final wave functions exactly coincides with the corresponding 
probability amplitude ${\cal A}_{if}$. In order to respect this below we 
shall designate the overlap integral of the initial and final wave 
functions (or quantum states) by using this notation. Note that $\hat{A}_{N}$ 
is a projection operator, then we can write the following identity 
$\hat{A}^{2}_{N} = \hat{A}_{N}$, which substantially simplifies the 
numerical calculations of the overlap integral, Eq.(\ref{overl}), since in 
reality we need only one $\hat{A}_{N}$ operator. Other useful properties of 
the overlap integral, Eq.(\ref{overl}), are mentioned, e.g., in 
\cite{Fro2010}. 

In general, it can be shown that the total final state probability for any 
nuclear $\beta-$decay almost equals unity, but in reality it is always 
smaller than unity. The explanation of this fact is obvious, since in our 
discussion above we ignored the bound-free transitions of the atomic 
electrons. In reality, we have to take into account the crucial fact that 
one (or a few) atomic electrons bounded to the maternal atom may have 
zero-projection(s) on the bound states of the final ion. In reality, the 
total probability to observe additional ionization of the final ion is 
small, but not negligible. For instance, for the one-electron tritium 
${}^{3}$H (or T) atom the total probability of such an additional 
ionization equals 1 - 0.97372715332 = 0.0262728467 (see, Table 1 in 
\cite{Fro2016}), i.e., $\approx$ 2.627285 \% of all $\beta^{-}$ decays 
in the tritium atom \cite{tritium} leads to the formation of completely 
ionized (or bare) ${}^{3}$He$^{2+}$ nucleus. 

In our previous papers on $\beta^{\pm}$ decays such a process of additional 
atomic ionization has never been investigated. In reality, the nuclear 
$\beta^{\pm}-$decays in atoms with an additional atomic ionization can be 
considered as an atomic bound-free transition which includes the unbound 
atomic states in the final atom/ion. It is clear that for such unbound 
states we cannot apply the same formulas such as Eq.(\ref{overl}) and 
Eq.(\ref{psiphi}), which were used above to determine the final state 
probabilities for the bound-bound transitions. However, the sudden 
approximation can still be used even in such cases, i.e., when the nuclear 
$\beta^{\pm}$ decay proceed with an additional atomic ionization. The only 
difference between bound-bound and bound-free atomic transitions is the 
correct choice of the final wave functions. This and a few other closely 
related problems are extensively discussed below and their solutions are 
the main goals of this study. 

\section{Nuclear $\beta^{-}-$decay with an additional ionization in a 
neutral few-electron atom} 

In general, the process of additional atomic ionization during the nuclear 
$\beta^{-}$ decay in some neutral atom $X$ is described by the equation 
\begin{equation}
\; \; \;  X(n L S) = X^{2+}(n^{\prime} L^{\prime} S^{\prime}) + e^{-} + 
\beta^{-} + \nu \; \; \label{betaA}  
\end{equation}
where the notation $\beta^{-}$ designates the fast $\beta-$electron, $\nu$ 
means neutrino, while the symbol $e^{-}$ denotes the regular (or slow) 
atomic electron emitted during the nuclear $\beta^{-}$ decay in the original 
(neutral) atom $X$. The notation $n L S$ designates the bound state in this 
atom. The emitted secondary (or slow) electron may have its own momentum 
$p$, angular momentum $\ell$ and spin ${\bf s}$ with the given orientations 
in our three-dimensional space. In fact, the projections of these two 
vectors on some spatial vectors can be measured at the final stage of the 
decay process, Eq.(\ref{betaA}). 

The main goal of this study is to derive the closed analytical formulas which 
can be used to evaluate (analytically and/or numerically) the final state 
probabilities of additional atomic ionization which is also possible during 
the nuclear $\beta^{-}$-decay in the few-electron atoms and ions. For these 
purposes we need to apply the explicit expression for the wave function of 
the unbound (or final) electron, which moves in the Coulomb field of the 
infinitely heavy, positively charged ion. It is clear that such a function is 
represented in the form $\phi_{e}({\bf r}) = \phi_{k \ell}(r) Y_{\ell m}({\bf 
n})$, where $\phi_{k \ell}(r)$ is the one-electron radial function, while 
$Y_{\ell m}({\bf n})$ is the corresponding spherical harmonic and ${\bf n} = 
\frac{{\bf r}}{r}$ is the unit vector associated with the radius-vector 
${\bf r}$. The parameter $k$ is the wave number which is uniformly related 
to the energy of the `free' electron $k = \sqrt{\frac{2 m_e E}{\hbar^2}} = 
\sqrt{2 E}$. Here and everywhere below in this study we apply the atomic 
units, where $\hbar = 1, m_e = 1, e = 1$ and $4 \pi \epsilon_{0}$ = 1. In 
these units the explicit formula for the single-electron radial part of the 
total wave function $\phi_{k \ell}(r)$ is written in the following form 
(see, e.g., Eq.(38.18) from \cite{LLQM}): 
\begin{equation}
 \phi_{k \ell}(r) = \frac{C_{k\ell}}{(2 \ell + 1)!} (2 k r)^{\ell} \; 
 \exp(-\imath k r) \; {}_1F_1\Bigl(\frac{\imath}{k} + \ell + 1, 2 \ell 
 + 2, 2 \imath k r \Bigr) \; \; \; \label{ubndwf}
\end{equation}
where $\imath$ is the imaginary unit, ${}_1F_1(a,b;x)$ is the confluent 
hypergeometric function (see, e.g., \cite{GR}), \cite{AS}) while 
$C_{k\ell}$ is the Coulomb penetration factor:
\begin{equation}
 C_{k \ell} = {\cal C}_{k 0} \; \prod^{l}_{s=1} \sqrt{s^2 + \frac{1}{k^2}} 
 = \sqrt{\frac{8 \pi k}{1 - \exp(-\frac{2 \pi}{k})}} \; \Bigl\{ 
 \prod^{\ell}_{s=1} \sqrt{s^2 + \frac{1}{k^2}} \Bigr\} \; \; . 
 \; \; \label{Cubndwf}
\end{equation}
This formula represents the `free' electron which moves in the central 
attractive field of the Coulomb potential which is created by some heavy 
bare nucleus and/or by a hydrogen-like ion. In general, the electric 
charge of central nucleus (or ion) is $Q e$ and its product with the 
electric charge of the electron $(- e)$ equals $- Q e^{2}$, or in atomic 
units this product equals - Q. However, if we do want to consider an 
atom (in atomic units) with arbitrary arbitrary (integer) charge of the 
nucleus $Q$, where $Q$ is integer and $Q \ge 1$, then in our formulas, 
Eqs.(\ref{ubndwf}) and (\ref{Cubndwf}) above, we have to make the  
following replacements: $k \rightarrow \frac{k}{Q}, r \rightarrow 
Q r$ and $k r \rightarrow k r$, i.e., the variable $k r$ does not change.
This concludes the first stage of our method. 

At the second stage we need to construct the radial part of 
single-electron wave function $\psi(r)$, which is included in the total 
wave function $\Psi(r)$ and describes to the original bound state in the 
maternal (or original) atom. Such a one-electron wave function can always 
be represented in the following Hylleraas-type form
\begin{equation}
 \mid \psi_{in}( r ) \rangle = \psi_{in}( r ) = \sum^{J_{max}}_{j=1} 
  {\cal B}_{j} \; r^{b_{j} - 1} \; \exp( - \lambda_{j} r) \; \; , 
  \; \; \label{Psi-in} 
\end{equation}
where ${\cal B}_{j}$ ($j = 1, 2, \ldots, J_{max}$) are the linear 
parameters of this representation, while the $b_{j}$ and $\lambda_{j}$ 
are the non-linear (or varied) parameters of this expansion. In many 
actual cases it is assumed that all $b_{j}$ are the positive integer 
numbers, while the parameters $\lambda_{j} = \lambda$ do not depend 
upon $j$. Nevertheless, even such a simple choice of non-linear 
parameters provides high numerical accuracy for the radial wave 
functions. Finally, the expression for the partial, single-electron 
transition amplitude $\langle \psi_{in}( r ) \mid  \phi_{k \ell}(r) 
\rangle$ takes the form
\begin{eqnarray}
 \langle \psi_{in}( r ) \mid  \phi_{k \ell}(r) \rangle = 
 \sum^{J_{max}}_{j=1} {\cal B}_{j} & & \frac{C_{k\ell} (2 
 k)^{\ell}}{(2 \ell + 1)!} \; \; \;\frac{\Gamma(\ell + b_{j} 
 + 2)}{\lambda_{j} - \imath k} \; \times \; \nonumber \\ 
 & & {}_{2}F_{1}\Bigl( 
 \ell + 1 - \frac{\imath}{k} , \ell + b_{j} + 2 , 2 \ell + 2 ; 
 \frac{2 k}{k + \imath \lambda_{j}} \Bigr) \; \; , \; \; \label{Aif}
\end{eqnarray}
where $\Gamma(x)$ is the Gamma-function, or Euler integral of the 
second kind. The two most important properties of the Gamma-function 
are $\Gamma(x + 1) = x \Gamma(x)$ and $\Gamma(1 - x) \Gamma(x) = 
\frac{\pi}{\sin x}$ (see, e.g., \cite{GR}, \cite{AS}). The function 
${}_{2}F_{1}( a, b; c; x)$ is the $(2,1)-$hypergeometric function 
defined exactly as in Eq.(9.100) from \cite{GR}. 

At the third stage of our method we need to derive the final formula 
for the probability amplitude which can be used for the nuclear 
$\beta^{-}-$decay with an additional ionization in the neutral 
$N-$electron atom. This many-electron part is crucial for workability 
of the whole method. Formally, the required expression directly 
follows from our formula Eq.(\ref{psiphi}) mentioned above where we 
have to make an obvious modification. Indeed, in Eq.(\ref{psiphi}) 
the last spin-orbital in the second determinant is simply replaced by 
the product of of radial and spin function $\chi_{N}$ of the final 
(or `free') electron. In other words, the actual replacement is 
written in the form: $\phi_{\beta_N}({\bf r}_{N}, {\bf s}_{N}) 
\rightarrow \phi_{k \ell}(r_{N}) Y_{\ell m}({\bf n}_{N}) \chi_{N}$. 
Numerical calculations of the arising overlap integral are not 
difficult and they allow us to determine the both final state 
probability amplitude and the final state probability. 

\section{Nuclear $\beta^{+}-$decay in a neutral few-electron atom}

The method developed above cannot be applied in those cases when the 
nuclear $\beta^{+}-$decay proceeds in neutral atoms. The most interesting 
atom in which one can easily observe the nuclear $\beta^{+}-$decay is the 
atom of ${}^{22}$Na isotope. In general, the $\beta^{+}-$decay process in 
${}^{22}$Na atoms proceeds with 89\% probability. It is competing with the 
electron capture (or $K-$capture) which has overall probability $\approx$ 
9.85\%. The half-time of the ${}^{22}$Na isotope against the nuclear 
$\beta^{+}-$decay is $\approx$ 2.58 years \cite{HAdl} The arising 
negatively charged ${}^{22}$Ne$^{-}$ ion is unstable and instantly decays 
into one `free' electron and neutral ${}^{22}$Ne atom. In this Section we 
want to obtain the correct formulas which can be used to determine the 
final state probabilities in the case of nuclear $\beta^{+}-$decay proceeds 
in neutral atoms. In this Section we shall apply the same three stage 
approach which we used above. 

First, it is necessary to chose the correct single-electron wave function 
which describes the `free' electron. To achieve this goal we note that in 
this case the final `additionally' ionized electron (or `free' electron, 
for short) moves in the central field of a neutral atom. The correct wave 
function of the final unbound (or `free') electron is written in the form 
$\phi_{k \ell}(r) Y_{\ell m}({\bf n})$, where $\phi_{k \ell}(r)$ is the 
one-electron radial function, while $Y_{\ell m}({\bf n})$ is the 
corresponding spherical harmonic and ${\bf n} = \frac{{\bf r}}{r}$ is the 
unit vector associated with the radius-vector ${\bf r}$ (see, e.g., 
Eq.(33.10) from \cite{LLQM}). The one-electron radial function takes the 
following (familiar) form:
\begin{equation}
 \phi_{k \ell}(r) = 2 k j_{\ell}(k r) = \sqrt{\frac{2 \pi k}{r}} \; 
 J_{\ell + \frac12}(k r) \; \; \; , \; \; \; \label{ubndwf1}
\end{equation}
where $j_{\ell}(x)$ is the spherical Bessel function of order $\ell$, 
where $\ell = 0, 1, 2, \ldots$, is a non-negative integer, while 
$J_{k}(x)$ is the Bessel function of the first kind. The product of this 
radial function with the angular factor $Y_{\ell m}({\bf n})$ determines 
the spatial function of the final (or `free') electron. To obtain the 
total wave function of this electron we also need to multiply by the 
electron's spin factor $\chi$. In reality, the wave function of the 
`free' electron which is emitted during the nuclear $\beta^{+}-$decay in 
a neutral few-electron atom is represented in the form $\phi_{k \ell}(r) 
\; Y_{\ell m}({\bf n}) \; \chi = 2 k j_{\ell}(k r) \; Y_{\ell m}({\bf n}) 
\; \chi = \sqrt{\frac{2 \pi k}{r}} \; J_{\ell + \frac12}(k r) \; 
Y_{\ell m}({\bf n}) \; \chi$. In general, such a wave function might also 
include a few additional indexes and/or internal parameters.   

At the second stage we need to construct single-electron wave function of 
the bound state. In reality, such a bound state wave function is exactly 
the same which we used in the previous Section. This allows us to move 
directly to the third stage of our method, i.e., we can obtain the 
following overlap integral between two radial single-electron functions 
Eq.(\ref{ubndwf1}) and Eq.(\ref{Psi-in}):  
\begin{eqnarray}
 a_{i \rightarrow f} = \langle \phi_{k \ell}(r) \mid \psi_{in}(r) 
 \rangle &=& \int_{0}^{+\infty} \; r^{b_{j} - 1} \; \exp(- \lambda_{j} 
 r) \; r^{2} \; \sqrt{\frac{2 \pi k}{r}} \; J_{\ell+\frac12}(k r) \; 
 dr \nonumber \\
  &=& \sqrt{2 \pi k} \; \int_{0}^{+\infty} \; r^{b_{j} + \frac12} \; 
  \exp(- \lambda_{j} r) \; J_{\ell+\frac12}(k r) \; dr \; \; \; , \; 
  \; \; \label{int1}
\end{eqnarray}
which exactly coincides with the partial transition amplitude $a_{i 
\rightarrow f}$ between two one-electron wave functions. After a few 
steps of additional transformations, which include the use of Eq.(6.621) 
from \cite{GR}, one finds the following explicit expression for the 
partial transition amplitude $a_{i \rightarrow f}(j)$ for the bound-free 
atomic transition in one-electron approximation:  
\begin{equation}
 a_{i \rightarrow f}(j) = \frac{\sqrt{\pi} \; k^{\ell + 1} \; \Gamma( 
 \ell + b_j + 2)}{2^{\ell} \; \sqrt{(\lambda^{2}_{j} + k^{2})^{b_{j} 
 + \ell + 2}\; \Gamma( \ell + \frac32 \Bigr)}} \; \; {}_{2}F_{1}\Bigl( 
 \frac{\ell + 2 + b_{j}}{2} , \frac{\ell - b_{j}}{2} , \ell + \frac32 
 ; \frac{k^{2}}{\lambda^{2}_{j} + k^{2}} \Bigr) \; . \; \label{Aif1}
\end{equation} 
Note that in this case all arguments of the hypergeometric functions 
are real, and moreover, the entire right-hand side of Eq.(\ref{Aif1}) 
is also real. In order to determine the total transition amplitude 
$A_{i \rightarrow f}$ and total final state probability $P_{i 
\rightarrow f}$ we have to apply the same procedure which is described 
at the end of previous Section. Deviations from the method described 
above are not significant. A few additional and useful details can be 
found in the Appendix.

\section{Conclusions} 

We have investigated the bound-free atomic excitations during the 
nuclear $\beta^{\pm}$-decay in the few- and many-electron atoms. In 
nuclear $\beta^{\pm}$-decay in the few- and many-electron atoms. can 
produce an additional ionization of the final atom/ion. Our main goal 
in this study was to derive the closed analytical formulas which can 
be used to determine the total probability amplitudes $A_{i 
\rightarrow f}$ for the final atomic states. In this study we could 
bring an important piece of exact mathematical science and rigorous 
logic to the area of research where all previous evaluations have 
been performed `by eye' and with the use of approximate functions only. 

Also, we have tote that the general theory of sudden atomic excitations 
already became an important part of modern atomic physics. In reality, 
there are a large number of atomic processes which follow the nuclear 
$\beta^{\pm}-$decay and probabilities of such events can now be 
calculated quite accurately. However, the current absence of even 
simple experiments in this area significantly restricts any further 
progress in research. It is clear that a number of new, accurate 
experiments are needed in various atoms, ions and molecules, if we want 
to compare our data from accurate data with actual experimental results. 

\appendix
 \section{Systems of wave functions of the continuous spectrum}
\label{A} 

In this study we have derived the explicit formulas which can be used 
to determine the total probability amplitudes and the final state 
probabilities in the case of nuclear $\beta^{\pm}-$decay in few- and 
many-electron atoms. In our approach to describe  the final, or `free' 
electron arising and moving after the nuclear $\beta^{\pm}-$decay we 
apply electronic wave functions written in the forms of 
Eq.(\ref{ubndwf}) and Eq.(\ref{ubndwf1}), respectively. However, 
results of some experiments can be explained only, if we represent 
such electronic wave functions as sums of plane waves, or by using 
some other functions. Therefore, it is necessary to understand other 
choices of different functions to describe the actual states from 
continuous spectra of atomic systems. 

In Quantum Mechanics there are a few different group of systems which 
are of great interest for theoretical analysis of atomic systems. 
First, there is a group of bound state problems, e.g., for atomic 
atoms and ions. Second, there are numerous scattering problems known 
in atomic physics. In addition to this, in this study we consider the 
problem of atomic decay when the final (or `free') electron arises 
and then moves from the central heavy `atom' to the infinite . In 
general, the motion of this `free' electron will be disturb by the 
acting electromagnetic field of the remaining atomic ion (or atom) 
and such an additional interaction must be taken into account 
correctly. Our main goal here is to choose the wave functions of the 
final electron which provide the correct radial asymptotic at large 
electron-nucleus and electron-ion distances. 

Note that it would be nice to use for our current purposes a number 
of theoretical methods known from atomic scattering, but for decay 
problems in atoms we obviously have no beam of incoming particles, 
e.g., electrons. The outgoing electron arises in the decay process 
itself. In sudden approximation the following motion of such a 
`free' electron can easily be predicted. This fact is the main and 
crucial difference between electron scattering and atomic decay. In 
respect to our classification scheme all atomic decays are included 
in the third group. To solve the problems from this third group we 
need to develop some new theoretical and computational methods. It 
is clear that these methods are completely different from the 
methods used to solve the problems from the first and second group. 

In general, any atomic decay is described by the decaying 
amplitude which plays the same role as the scattering amplitude 
plays in atomic scattering. However, for atomic decays we have no 
interference between the beam of incoming electrons and secondary 
electrons which are formed after the nuclear $\beta^{\pm}-$decay. 
Briefly, this means that we have no optical theorem for atomic 
$\beta^{\pm}-$decays. Therefore, all relations and consequences 
of this theorem can be ignored for atomic decay problems. This 
substantially simplifies our analysis below. Further analysis 
indicates clearly that we need only the converging and diverging 
spherical and/or parabolic waves which are designated in modern 
literature as the $\psi^{-}_{{\bf k}}$ and $\psi^{+}_{{\bf k}}$ 
wave functions of the continuous atomic spectrum.  

The explicit forms of these wave functions are (see, e.g., 
\cite{LLQM}):  
\begin{eqnarray}
 \psi^{+}_{{\bf k}} &=& \frac{1}{2 k} \; \sum^{\infty}_{\ell=0} 
 \; \imath^{\ell} \; (2 \ell + 1) \; \exp(\imath \delta_{\ell}) 
 \; R_{k\ell}(r) \; P_{\ell}\Bigl(\frac{{\bf k r}}{k r}\Bigr) 
 \; \; \; , \; \; \; \label{psi+} \\
 \psi^{-}_{{\bf k}} &=& \frac{1}{2 k} \; \sum^{\infty}_{\ell=0} 
 \; \imath^{\ell} \; (2 \ell + 1) \; \exp(- \imath \delta_{\ell}) 
 \; R_{k\ell}(r) \; P_{\ell}\Bigl(\frac{{\bf k r}}{k r}\Bigr) 
 \; \; \; , \; \; \; \label{psi-} 
\end{eqnarray}
in the case when the negatively charged, `secondary' electron 
moves away from the neutral atom and ion. In these equations 
${\bf k}$ is the wave vector of this electron, $P_{\ell}$ are 
the Legendre polynomials of the following cosine function: 
$\cos\theta = \Bigl(\frac{\bf k r}{k r}\Bigr)$. The notations 
$R_{k\ell}(r)$ designate the system of radial functions which 
are exactly coincides with the spherical Bessel functions of 
order $\ell$ defined in Eq.(\ref{ubndwf1}). 

Note that the wave functions Eq.(\ref{psi+}) and Eq.(\ref{psi-}) 
obey the two following equations:
\begin{eqnarray}
 \int \Bigl(\psi^{+}_{{\bf k}_1}\Bigr)^{*} \psi^{-}_{\bf k}
 dV = (2 \pi)^{3} \delta({{\bf k}_1} - {\bf k}) \; \; \; {\rm 
 and} \; \; \; \; \psi^{-}_{{\bf k}} = \Bigl(\psi^{+}_{{\bf 
 -k}}\Bigr)^{*} ; \; \; , \; \; \; \label{psirpo} 
\end{eqnarray}
where the sign $a^{*}$ means complex conjugate of $a$. These 
equations are truly important to show the completeness of this 
systems of wave functions and simplify their applications to 
various actual problem. As was shown by A. Sommerfeld in 1931 
\cite{Somm} in real calculations the transition probability of 
any actual process one needs to apply the converging spherical 
function only, i.e., the $\psi^{-}_{\bf k}$ function, or its 
complex conjugate $\Bigl(\psi^{-}_{\bf k}\Bigr)^{*}$ to be 
exact.   

To conclude this Section we need to discuss the case when the 
negatively charged, `secondary' electron moves away from the 
positively charged ion. This case describes more than 99\% 
of all actual atomic $\beta^{-}-$decays. In this case our 
wave functions from Eq.(\ref{psi+}) and Eq.(\ref{psi-}) cannot 
be used and we have to replace them with some other wave 
functions which are more accurate for description of the 
Coulomb attractive potential. In atomic units the explicit 
formulas for these wave functions are (see, e.g., \cite{BetMax}  
and \cite{LLQM}): 
\begin{eqnarray}
 \psi^{+}_{{\bf k}} &=& \exp\Bigl(\frac{\pi Q}{2 k}\Bigr) \; 
 \Gamma\Bigl( 1 - \frac{\imath Q}{k}\Bigr) \; \exp( \imath 
 {\bf k r}) \; \; {}_{1}F_{1}\Bigl( \frac{\imath Q}{k} , 1 ; 
 \imath ( k r - {\bf k r} )\Bigr) 
 \; \; \; , \; \; \; \label{Cpsi+} \\
 \psi^{-}_{{\bf k}} &=& \exp\Bigl(\frac{\pi Q}{2 k}\Bigr) \; 
 \Gamma\Bigl( 1 + \frac{\imath Q}{k}\Bigr) \; \exp( \imath 
 {\bf k r}) \; \; {}_{1}F_{1}\Bigl( -\frac{\imath Q}{k} , 1 ; 
 - \imath ( k r - {\bf k r} )\Bigr) 
 \; \; \; , \; \; \; \label{Cpsi-}  
\end{eqnarray}
where $\Gamma(x)$ is the Euler $\Gamma-$function and ${}_{1}F_{1}( 
a, b; x)$ is the confluent hypergeometric function. Sometimes these 
wave functions are called the parabolic waves, since the variable 
$k r - {\bf k r} = z$ in these equations coincides with the 
parabolic coordinate $z$. It is clear that each of the parabolic 
wave functions has only azimuthal (not spherical) symmetry. These 
two formulas, Eqs.(\ref{Cpsi+}) - (\ref{Cpsi-}), solve our problem  
formulated above. In general, the wave functions, Eq.(\ref{Cpsi+}) 
and Eq.(\ref{Cpsi-}) provide the correct asymptotics (at large 
distances) in the Coulomb field and in the non-relativistic 
approximation. Various integrals between these two functions and 
partial derivatives of such integrals upon different parameters 
play important role in Quantum Electrodynamics (see, e.g., 
\cite{AB}), since they are included in answers of many actual QED 
problems.


\begin{thebibliography}{01}

\bibitem{Blat} J.M. Blatt and V.F. Weisskopf, \textit{Theoretical Nuclear 
Physics}, (Springer-Verlag Inc., New York (1979)). 

\bibitem{Mig1} A.B. Migdal, J. Phys. (USSR) 4, 449 (1941). 

\bibitem{Mig2} A.B. Migdal and V. Krainov, {\it Approximation Methods 
in Quantum Mechanics}, (W.A. Benjamin, New York (1969)). 

\bibitem{Fro2005} A.M. Frolov and J.D. Talman, Phys. Rev. A {\bf 72}, 
022511 (2005).

\bibitem{Fro2010} A.M. Frolov and M.B. Ruiz, Phys. Rev. A {\bf 82}, 
042511 (2010). 

\bibitem{Fro2016} A.M. Frolov, Letters to JETP, {\bf 103}, 835 (2016) 
[Russian Trasl.: JETP Letters, {\bf 103}, 739 (2016)]. 

\bibitem{McWen} R. McWeeny and B.T. Sutcliffe, {\it Methods of Molecular
Quantum Mechanics}, (Academic Press, New York (1969)), Chapters III and 
IV. 

\bibitem{tritium} The tritium atoms play an important role in the 
development of modern theory of atomic excitations during nuclear 
$\beta^{\pm}$-decays. The existence and instability of tritium atoms was 
discovered in 1934 after the collisions of two accelerated deuterium 
nuclei. In 1939 Alvarez and Cornog found that tritium decays by the 
nuclear $\beta^{-}$-decay. Later it was found that the half-life of 
tritium nuclei against such a decay is $\approx$ 12.262 years, while the 
maximal energy of the emitted $\beta^{-}$ particles do not exceed 18.3 
$keV$. 

\bibitem{LLQM} L.D. Landau and E.M. Lifshitz, {\it Quantum Mechanics: 
non-relativistic theory}, (3rd. ed. Pergamon Press, New York (1976)), 
Chpt. VI. 

\bibitem{GR} I.S. Gradstein and I.M. Ryzhik, \textit{Tables of Integrals, 
Series and Products}, (6th revised ed., Academic Press, New York (2000)).

\bibitem{AS} \textit{Handbook of Mathematical Functions} (M. Abramowitz 
and I.A. Stegun (Eds.), Dover, New York, (1972)).

\bibitem{HAdl} H. Haissinsky and J.-P. Adloff, {\it Radiochemical Survey 
of the Elements. Principal Characteristics and Applications of the 
Elements and their Isotopes}, (Elsevier Publ. Comp., Amstredam (1965)). 

\bibitem{Somm} A. Sommerfeld, Atombau und Spektrallinien: Vol. 1. (5th 
Edition, Vieweg, Braunschweig, (1931)).

\bibitem{BetMax} H.A. Bethe and L. Maximon, Phys. Rev. {\bf 93}, 768 (1954). 

\bibitem{AB} A. Akhiezer and V.B. Berestetskii, \textit{Quantum 
Electrodynamics} (4th ed., Science, Moscow (1981)) [in Russian], 
Section 4.3.3. 

\end{thebibliography}
\end{document}